\documentstyle[aps,pre,psfig,epsfig,twocolumn]{revtex}
\begin{document}
\bibliographystyle{unsrt}
\draft
\title{Soliton ratchets induced by ac forces with harmonic mixing}
\author{Mario Salerno$^\dag$ and Yaroslav Zolotaryuk$^\ddag$}
\address{
$^\dag$ Dipartamento di Fisica ``E.~R.~Caianiello'' and  Istituto
Nazionale di
 Fisica della Materia (INFM),\\ Universit\'{a} di Salerno, I-84081 Baronissi,
Salerno, Italy \\
$^\ddag$ Section of Mathematical Physics, IMM, Technical
University of Denmark, \\DK-2800, Kgs. Lyngby, Denmark }
\date{\today}
\wideabs{
\maketitle

\begin{abstract}
The ratchet dynamics of a kink (topological soliton) of a
dissipative sine-Gordon equation in the presence of ac forces with
harmonic mixing (at least bi-harmonic) of zero mean is studied.
The dependence of the kink mean velocity on system parameters is
investigated numerically and the results are compared with a
perturbation analysis  based on a point particle representation of
the soliton. We find that first order perturbative calculations
lead to incomplete descriptions, due to the important role played
by the soliton-phonon interaction in establishing the phenomenon.
The role played by the temporal symmetry of the system in
establishing soliton ratchets is also emphasized. In particular,
we show the existence of an asymmetric internal mode on the kink
profile which couples to the kink translational mode through the
damping in the system. Effective soliton transport is achieved
when the internal mode and the external force get phase locked. We
find that for kinks driven by bi-harmonic drivers consisting of
the superposition of a fundamental driver with its first odd
harmonic, the transport arises only due to this {\it internal
mode} mechanism, while for bi-harmonic drivers with even harmonic
superposition, also a point-particle contribution to the drift
velocity is present. The phenomenon is robust enough to survive
the presence of thermal noise in the system and can lead to
several interesting physical applications.
\end{abstract}
}

Transport phenomena based on nonlinear effects are at the heart of
many problems in physics \cite{suz92}. In this context it was
generally believed that an ac force of zero mean cannot lead to
directed nonzero currents. Recent studies of the so called  {\it
ratchet effect} have shown that this belief was wrong
\cite{hb96-jap97}. A ratchet system can be described as a Brownian
particle in an asymmetric  periodic potential,  moving in a
specific direction in presence of damping, under the action of ac
forces of zero average. The origin of a net motion is associated
to the breaking of the space-temporal symmetries of the system
\cite{fyz00prl,yfzo01epl}, leading to the de-symmetrization of
L\'evy flights [for Hamiltonian systems] \cite{df01} and to phase
locking phenomena between the particle motion and the external
driving force \cite{barbi,barbi-sa}.

This effect has a number of applications in various branches of
physics and biology and is believed to be the basic mechanism for
the functioning of biological motors (see reviews
\cite{hb96-jap97} and references therein). The ratchet effect,
originally studied for Brownian particles, was generalized to
dynamical systems \cite{jkh96prl} and to partial differential
equations (PDE) of soliton type, mainly in the over-damped regime
\cite{m96prl}, or with asymmetric potentials
\cite{stz97pla,gsk01pre,fmmc99epl,sq01,cm01prl}. In the over-damped
case, the damping in the system suppress all the degrees of
freedom associated with the background radiation so that soliton
ratchets become very similar to point particle ratchets. For
under-damped or moderately damped systems, however, the situation
is quite different since the radiation field can interact with the
soliton and influence the transport. Under-damped soliton ratchets
in asymmetric potentials and driven by sinusoidal forces, were
recently investigated in Refs. \cite{sq01,cm01prl}. In particular,
in Ref. \cite{sq01} the basic mechanism underlying the phenomenon
was identified in the existence of an asymmetric internal mode
which couples, through the damping in the system, to the soliton
translational mode. Effective transport was found when the
internal mode and the external force were phase locked. Moreover,
it was shown that the effect of soliton transport decreases with
increase of the damping, the maximal transport being achieved in
the under-damped regime.

On the other hand, it is known that for point particle ratchets
transport phenomena are possible also in symmetric  periodic
potentials, provided the driving force breaks suitable temporal
symmetries of the system (i.e. the ones which relates orbits of
opposite velocities in phase space \cite{mix}). Since in concrete
applications it is more easy to act on the temporal part (by using
external forces) than on the spatial part (by inducing distortion
of the potential) of a system, it is interesting to explore the
above possibility also in soliton systems.

The present paper  is just devoted to this problem. More
precisely, we show that topological solitons of nonlinear PDEs
with symmetric potentials can acquire finite drift velocities in
the presence of bi-harmonic forces of zero average consisting of
the superposition of two harmonics, the fundamental and one of its
overtones (harmonic mixing drivers). Bi-harmonic forces were used
in the literature to suppress chaos in dynamical systems and in
soliton equations \cite{salerno}, as well as, to control the
transport properties of single particle ratchets \cite{barbi-sa}.
In the present paper we demonstrate, on the particular example of
the sine-Gordon system, that bi-harmonic driving forces with
certain symmetries can be effective to create soliton ratchets.
The phenomenon is investigated both numerically, by direct
simulations, and analytically, using symmetry considerations and
soliton perturbation theory. We show that a first order
perturbation analysis of the soliton dynamics captures only the
qualitative features of the phenomenon, leading to poor
quantitative agreements with numerical results. The reason of this
discrepancy is ascribed to the soliton-phonon interaction which is
obviously missing in a point-particle description (it arises only
at the second order in the perturbation expansion \cite{ss82}).

In general the situation can be described as follows. Besides a
point particle contribution to the drift velocity there is an
equally important contribution coming from the soliton-phonon
interaction. This last manifests itself with the appearance of an
internal oscillation on the soliton profile, asymmetric in space,
which induces a net motion in a similar manner  as described in
Ref. \cite{sq01}. In particular we show that this oscillation can
be phase locked to the external driving force and can couple to
the kink translational mode, trough the damping in the system. The
energy, ``pumped'', by the ac field into the internal mode is then
converted by the above mechanism into net motion of the kink.
Internal oscillations on anti-kink profiles have opposite
asymmetry compared to kinks, so that kink and anti-kink ratchets
give rise to motion in opposite directions.

The dependence of the phenomenon on system parameters such as the
damping in the system, frequency, amplitudes, relative phase of
the bi-harmonic driving force, as well as, on the presence of
white noise in the system, is investigated by means of direct
numerical integrations of the sine-Gordon equation. The
interaction of soliton ratchets with the boundaries of a finite
system is also investigated. For reflecting edges we find that,
depending on the initial velocity and position of the kink, the
ratchet dynamics can be either destroyed or reflected as an
anti-kink ratchet moving in the opposite direction. These results
could be important for applications to physical systems such as
long Josephson junctions, as we briefly discuss at the end of the
paper.

The paper is organized as follows. In Section I we investigate the
dynamics of the perturbed sine-Gordon equation, as a model for
soliton ratchets in presence of asymmetric forcing and damping,
both in terms of symmetry arguments and first order perturbation
theory. In Section II we study soliton ratchets by direct
numerical integrations of the perturbed sine-Gordon system and
provide a consistent interpretation of the phenomenon. Qualitative
and quantitative features of soliton ratchets are compared with
the predictions of the first order perturbation analysis. The
phenomenon is investigated in the deterministic case (i.e. in
absence of noise) and in presence of a white noise term in the
system. We find  that the soliton ratchets are robust enough to
overcome the presence of small amplitude noises, this making them
of interest for practical applications. In Section III  the
interaction of soliton ratchets with system  boundaries is
considered, while in Section IV we summarize the main result of
the paper and discuss possible applications of the phenomenon.

\section{Model Analysis}

Direct soliton motion induced by ac signals have been investigated
in the literature mainly for ac forces with single-harmonic
content \cite{km89rmp}. As well known, for symmetric field
potentials this situation does not lead to soliton ratchet
dynamics. To this regard we remark that the dc motion observed for
a sine-Gordon kink driven by single harmonic forces in absence of
damping \cite{os83prb}, and its generalization to the case of
small damping \cite{qs98epj}, as well as, dc motion obtained from
spatially-inhomogeneous drivers \cite{malomed}, should not be
confused with soliton ratchets. In these cases, indeed, the dc
motion strongly depends on the initial conditions and quickly
disappears as the damping in the system is increased. On the
contrary, soliton ratchets do not depend on initial conditions and
exist also for relatively higher damping. We remark that net
soliton motion independent on initial conditions, can be induced
by the mixing of an additive and a parametric (single harmonic)
driver as shown in Ref. \cite{sp95prl} for solitons of the
$\phi^4$ model.

In this section we shall investigate soliton ratchets in symmetric
potentials driven by periodic bi-harmonic forces of zero mean. As
a  working model we take  the following perturbed sine-Gordon
equation
\begin{equation}
u_{tt} - u_{xx} + \sin {u}=-\alpha u_t - E(t) + n(x,t)\;\;,
\label{1}
\end{equation}
with $\alpha$ denoting the damping coefficient, $n(x,t)$  a white
noise term with autocorrelation
\begin{equation}
<n(x,t){n}(x^{\prime },t^{\prime })>=D\delta (x-x^{\prime })\delta
(t-t^{\prime }),
\label{noise0}
\end{equation}
and $E(t)$  a driver of the form
\begin{equation}
E(t) =E_1 \cos {\omega t} + E_2 \cos {(m \omega t+\theta)}\;\;,
\label{2}
\end{equation}
(the case $E_2\neq 0$ is referred to as {\it bi-harmonic driver}
with even or odd harmonic mixing, depending on $m$ being even or
odd). Note that although the symmetry properties of this driver
are reduced if $E_2 \neq 0$ and $\theta \neq 0$ $\mbox{mod}~\pi$,
the force is periodic, with period $T=2\pi/\omega$, and has zero
mean (the analysis can be generalized to more harmonic components
and to arbitrary non-sinusoidal periodic forces). The unperturbed
version of Eq. (\ref{1}) [the perturbation being $\epsilon
f(t)\equiv -E(t)-\alpha u_t(x,t)$] is the well known sine-Gordon
equation with exact soliton (kinks, anti-kinks) solutions  which
depend on a free parameter, the velocity $v$ of the kink, which
lies in the range $-1<v<1$.

It is of interest to investigate the conditions under which a
bi-harmonic driver of type (\ref{2}) can induce soliton ratchets
in Eq. (\ref{1}). To this end we remark that due to the
translational invariance of the sine-Gordon system (we assume an
infinite system  or a finite one with periodic boundary
conditions) we have that to each soliton trajectory with velocity
$v$ there is a specular trajectory with the velocity $-v$. One can
expect that, in analogy with single particle ratchets \cite{mix},
only forces which break the $v \rightarrow -v$ symmetry should
induce net motion. This argument can be formalized in terms of the
kink velocity
\begin{equation}
v=\frac{1}{2\pi} \int_{-\infty}^{+\infty} x u_{xt} dx~,
\end{equation}
as follows (note that one could use the momentum of the kink as
well,  instead of the velocity).  Among the possible shifts and
reflections in $t,x$ and $u$, we identify the symmetry operations
which change the sign of $v$ keeping the sign of the topological
charge
\begin{equation}
Q=\frac{1}{2\pi}\int_{-\infty}^{+\infty} u_{x} dx~,
\end{equation}
unchanged (this means that we avoid kink anti-kink
transformations). It is easy to check that there is only one
symmetry transformation which changes the sign of kink velocity
and leaves the equation of motion unchanged, i.e.
\begin{equation}
x \rightarrow  -x + x_0~,~~ t \rightarrow t+ \frac{T}{2}~,~~ u
\rightarrow - u + 2\pi~.
\end{equation}
This holds true for driving fields satisfying the following
condition
\begin{equation}
E(t+T/2)=-E(t)~. \label{7}
\end{equation}
We remark that the above symmetry argument accounts only for the
breakage of the $v \rightarrow -v$ point-particle symmetry of the
soliton, ignoring possible contributions coming from the
soliton-phonon interaction \cite{flach}. From Eq. (\ref{2}) we see
that condition (\ref{7}) is always satisfied for drivers with odd
mixing while it is always broken for drivers with $m$ even
(obviously we take $E_2 \neq 0$). Thus, the $v \rightarrow -v$
symmetry predicts that a sine-Gordon soliton should exhibit a
ratchet dynamics when driven by a $m=2$ force  (one needs to break
the symmetry (\ref{7}) to get the drift motion), but not when
driven by a $m=3$ force.

To the same conclusion one can arrive also from first order
perturbation theory, taking as collective coordinate ansatz for
the kink profile
\begin{equation}
u(x,t)=4 \arctan \left ( \exp \left[ \frac{x-X(t)} {\sqrt{1-\dot
{X}^2(t)}}\right ] \right )~.
\label{4bis}
\end{equation}
An ordinary differential equation (ODE) for kink's center of mass
$X(t)$, can be readily obtained by differentiating the momentum
(see Refs. \cite{os83prb,mls78pra}),
\begin{equation}
\label{5bis}  P=-\int_{-\infty}^{+\infty} \phi_x \phi_t dx~,
\end{equation}
with respect to time, and using Eq. (\ref{1}) and ansatz
(\ref{4bis}) to simplify the expression. This leads to
\begin{equation}
\frac{dP}{dt}=- \alpha P + 2 \pi E(t)~,
\label{eqmom}
\end{equation}
or equivalently
\begin{equation}
\dot {v}=-\frac{1}{4}(1-v^2)\left [-\pi \sqrt{1-v^2}E(t)+4\alpha v
\right],~
\label{6}
\end{equation}
where $v(t)=\dot X(t)$ and we assumed the usual relativistic
relation $P(t)=8 v(t)/\sqrt{1-v(t)^2}$  between velocity and
momentum to be valid for all times. Equation (\ref{eqmom}) can be
readily solved for $P(t)$, from which  the kink velocity can be
obtained as
\begin{equation}
\label{velocity} v=\frac{P(t)}{8\sqrt{1+\frac{P^2(t)}{64}}}~.
\end{equation}
From this equation an analytical expression of the average kink
velocity
\begin{equation}
\langle v \rangle =\frac{\omega}{2 \pi}
\int_0 ^T v(t') dt'~,~~T=\frac{2 \pi}{\omega}~,
\label{6a}
\end{equation}
valid in the limit $E_j/\sqrt{\alpha^2+ (j \omega)^2}<<1$,
$j=1,2$, (i.e. small momentums or small drift velocities), can be
obtained for the case of $m=2$, by expanding the square root in
Eq. (\ref{velocity}) in series, this giving
\begin{eqnarray}
\label{v-approx} \langle v \rangle &=& \frac{3}{512}\frac{E_1^2
E_2 \pi^3}{(\alpha^2+\omega^2)\sqrt{\alpha^2+4\omega^2}}
\sin(\theta-\theta_0)~,\\
\theta_0 &=& \arctan \left [\frac{\alpha (\alpha^2+3\omega^2)}{2
\omega^3} \right ]~. \nonumber
\end{eqnarray}
From this expression we see that the dependence of $\langle v
\rangle$ on the relative phase is perfectly sinusoidal. Similar
calculations for the $m=3$ case show that the average kink
velocity is zero independently on the value of the relative phase
$\theta$ (as well as of an arbitrary initial phase).

These results can be easily understood from the effects of the
symmetry properties of the force on the dynamics. In Fig.
\ref{fig0ms} the force $E(t)$, viewed as sequence of alternating
pulses of equal intensities (i.e. with the same area under the
curve) and indicated by dark and light fillings in the figure, is
reported for the cases $m=2,3$. We see that for $m=3$ these pulses
perfectly balance so that no net motion can arise, while for $m=2$
there is not such a balance (in both cases, however, the average
of the force is zero).
%

In Fig. \ref{fig3ms} we report the dynamics obtained from
numerical integrations of Eq. (\ref{6}) for the case $m=2$. We
also show the response of the system to the single dark and light
pulses composing the force, from which we see that although these
pulses have equal intensities, the answer of the system is quite
different in the two cases. Note that the negative pulse is more
effective than the positive one to produce motion as one can see
from the fact that the area under the negative trajectory is
greater than the one generated by the positive pulse. This is
obviously a consequence of the nonlinearity of the system (in a
linear system the area under the two curves would just be the
same). For the particular example of Fig. \ref{fig3ms} (i.e.
$\theta=0$) one expects then that a net motion in the negative
direction will exist . This is indeed what one finds from
integrations of the perturbation equation (\ref{6}) with
$\theta=0$ [see Fig. (\ref{fig2ms}) below].
%

In Fig. \ref{fig1ms} the motion of the kink center of mass, $X$,
obtained from numerical simulations of Eq. (\ref{6}) for the cases
$m=2,3$, is also reported. We see that while a well defined drift
velocity in the $m=2$ case arises, no dc motion is present in the
$m=3$ case, this being in agreement with our symmetry analysis.

We also find that the average velocity in (\ref{6a}), computed
after the system reached steady state regime, depends on the
relative phase $\theta$ with a low which is be well approximated
by $\langle v \rangle= A \sin (\theta - \theta_0)$, with $A=0.058$
and  $\theta_0=0.8$. This is shown in Fig. \ref{fig2ms} where the
dependence of the average velocity on $\theta$, as computed from
Eq.(\ref{6}), is reported with dots, while the continuous line
represents the above approximating function. Note that, although
the sinusoidal dependence is in perfect agreement with the
approximate result in Eq. (\ref{v-approx}), the explicit values of
$A, \theta_0$ differ from those predicted by (\ref{v-approx}),
these being respectively $A=0.12$, $\theta_0=0.56$ (this is due to
the fact that, for the chosen set of parameters, the approximation
of small drift velocities is not valid).

%

%

We also note that from Eq. (\ref{v-approx}) the point-particle
contribution to the drift velocity is expected to be cubic in the
driver amplitudes, this implying that there could be equally
important contributions to the soliton velocity at higher orders
of the perturbation expansion. In particular, the interaction of
the soliton with the phonons in the system, first appearing at
second order \cite{ss82}, should not be overlooked. Although the
development of a theory which includes second order effects is
quite challenging, it is out of the purposes of the present paper.
In the next section will shall instead resort to numerical
simulations of Eq. (\ref{1}) for a full investigation of the
problem  and compare the results with those predicted by the
present section.

\section{Numerical studies}

To numerically investigate sine-Gordon soliton ratchets driven by
bi-harmonic fields, we have used standard finite difference
schemes to reduce Eq. (\ref{1}) into a set of ODE which were then
integrated in time with a 4th order Runge-Kutta method. To
understand the basic mechanism underlying the phenomenon we shall
first concentrate on the deterministic case by putting $n(x,t)=0$
in Eq. (\ref{1}), and then show that the results obtained in this
case will survive in the presence of noise.
%
In Fig. \ref{fig1} the dynamics of a sine-Gordon kink, initially
at rest, driven by a bi-harmonic driver with $m=2$  and by a
single harmonic driver (i.e. $E_2=0$), are reported in Figs.
\ref{fig1}a,b, respectively (note that we use contour plots to
show the time evolution surface generated by the kink profile).
From these figures we see that, in the case $m=2$ the soliton
center of mass move with a constant drift velocity (note that the
shape of the kink during the motion is highly distorted), while
for the single harmonic driver, it oscillates around the initial
position (periodic boundary conditions are used in the
simulation). This demonstrates the importance of bi-harmonic
drivers in establishing soliton ratchets.

In Fig. \ref{fig2} we report the dynamics of an anti-kink ratchet,
for the same parameters values as in Fig. \ref{fig1}. We see that,
in analogy with soliton ratchets in spatially asymmetric
potentials \cite{m96prl,stz97pla,sq01}, anti-kinks ratchets move
opposite to kinks, the absolute value of the drift velocity being
the same.

To check the point particle perturbation analysis of the previous
section, we have studied the dependence of kink's average velocity
on system parameters for the case of a bi-harmonic driver with
$m=2$.
%
%

In Fig. \ref{fig3} we report $\langle v \rangle$ as a function of
$\alpha$  for different values of system parameters. Note that the
curves display similar behaviors, with a resonance peak in the
under-damped regime and a quick decay to zero at higher damping.
This behavior is similar to the one reported in Ref. \cite{sq01}
for soliton ratchets in asymmetric potentials, and suggests a
possible common mechanism of the phenomenon (see below). We also
remark that the interruptions of the curves (cut-offs) at small
$\alpha$ values are due to the disappearance of the kink as a
consequence of the onset of spatio-temporal chaos in the system
(the cut-offs delimit the borders of the existence diagram of the
kinks).

Panel (a) of Fig. \ref{fig3} shows the dependence of $\langle v
\rangle$ on $\alpha$ for different values of the driving
frequency. Note that the velocity is influenced by resonances with
the plasma frequency and its harmonics, as one can see from the
non-monotonous behavior of the cut-offs at small $\alpha$ (the
cut-off velocities in this case have their largest values close to
$\omega=0.5$ and $\omega=1$).

%

Another interesting property emerging from this figure is that the
kink velocity is enhanced at low frequencies, and the average
velocity decreases by increasing the frequency (already for
frequencies above $\omega=1$ directed motion is hardly visible).
This is a consequence of the kink inertia to react to fast
oscillations. On the contrary drift velocities are observable also
at quite small values of $\omega$ (at very small values, however,
the dynamics becomes complicated and requires long computational
times - we checked explicitly the case $\omega=0.01$ for which an
average drift is still visible). The dashed lines in the figure
represent the results of the point particle perturbation analysis
of the previous section. We see that the agreement, although
qualitatively reasonable, is not so good from a quantitative point
of view.

In panel (b) of Fig. \ref{fig3} the dependence of $\langle v
\rangle$ on $\alpha$ is reported for different driver amplitudes
(for simplicity  we have varied $E_1$, fixing the ratio
$E_2/E_1=0.65$ ). We observe a situation similar to the one shown
in Fig. \ref{fig3}a. By increasing the driver amplitude the system
reaches the chaotic regime and  cut-off values in $\alpha$ quickly
appear. The resonant peak in this case is quite weak and visible
only for some narrow range of driving amplitudes. Also here the
predictions of the point particle perturbation theory are
quantitatively quite poor. In the panel (c) of Fig. \ref{fig3} we
have shown the dependence $\langle v \rangle(\alpha)$ for
different values of the relative phase $\theta$. We see that by
changing $\theta$ one can change a maximum of the curve at a given
value of $\alpha$, into a minimum. This is a consequence of the
sinusoidal dependence of the average velocity on $\theta$
predicted in the previous section. To show this, we have reported
in the inset of the figure $\langle v \rangle$ versus $\theta$ for
a fixed value of $\alpha$, as computed from direct numerical
integrations of the sine-Gordon system. We see that the numerical
points are well fitted by a sinusoidal law as expected from the
first order perturbation result of the previous section. We remark
that a similar sinusoidal dependence was also found in Ref.
\cite{yfzo01epl} for single ODE ratchets, this confirming the
existence of a point-particle contribution to the effect.
%
In experimental situations in which the relative phase between the
two drivers is not accessible, one should consider $\theta$ as a
random variable, and a final average on it should be taken. The
above results then imply that no drift velocity can exist in these
cases (soliton ratchets can be induced only if the relative phase
$\theta$ remains constant in time).

It is also interesting to note from Fig. \ref{fig3}c that reversal
currents can be induced by changing the relative phase. In Fig.
\ref{fig4} we show  how the curve $\theta=\pi$ of Fig.
\ref{fig3}c, (which display current reversal at low damping),
changes as the driving amplitude is increased. We see that by
increasing the amplitude of the driver the kink velocity is
increased, this leading to an upwards shift of the curve. This
means that current reversal observed for some value of $\theta$
can be removed by properly adjusting the driving amplitude (and
vice-versa). Note that a further increase of the amplitude can
change the shape of the curve destroying the resonant-like
character.

In Fig. \ref{fig10} we report the dependence of the average
velocity on the driver amplitude $E_1$ with the ratio $E_2/E_1$
fixed to $0.65$, and with the relative phase between the two
drivers $\theta=1.61$. From this figure it is clear  that the kink
drift velocity depends non linearly on the driver amplitude $E_1$,
with an almost cubic low as one can see from the $log-log$ plot in
the inset. This result is in good qualitative agreement with our
perturbation analysis (note that the $E_1^2 E_2$ dependence in Eq.
(\ref{v-approx}) implies a $E_1^3$ dependence if one fixes the
ratio $E_1/E_2$, as done in the numerical simulations). We remark,
however, that the case $\omega=0.11$, denoted by $*$ in Fig.
\ref{fig10}, indicates a deviation from this law at higher values
of $E_1$.
%

We also  checked the predictions of point particle symmetry
arguments and perturbation theory for the case of bi-harmonic
forces with odd harmonic mixing. In Fig. \ref{k3} the dynamics of
a kink driven by a bi-harmonic force with $m=3$ is reported. In
contrast with the prediction of the previous section, we see that
kink can acquire a drift velocity also in this case. The direction
of the motion, however, depends not only on the relative phase,
but also on the initial phase (or initial time) of the driving
force. By changing the initial phase  we can achieve soliton
moving  in the opposite direction with the same velocity, thus by
averaging on the initial phase one gets a zero mean velocity for
the kink [simulations of the kink dynamics with different initial
phases in the interval $[0,T[$ show that roughly half of that
interval of initial points yield attraction to kink solutions
moving to the right, the other half leading to kink solutions
moving to the left with the same velocity]. A similar study for
the case of $m=2$ did not show any dependence on the initial phase
or on initial time. Due to the sensitivity on initial conditions,
we conclude that the kink motion in the case $m=3$ is not related
to the ratchet phenomenon  (this is similar to the cases reported
in Refs.\cite{os83prb,qs98epj}).

The net motion observed for the $m=3$ case for fixed values of the
initial phase, however, is an interesting phenomenon to explore by
itself, since it is not linked with point-particle features of the
soliton dynamics (these are excluded by the results of the first
order perturbation theory), and is entirely related to the
soliton-phonon interaction in a similar way as discussed in Ref.
\cite{sq01}.

%

From the above analysis the following conclusions can be drawn.
Although first order perturbation theory captures some qualitative
feature of soliton ratchets, it does not provide a satisfactory
description of the phenomenon. This is clear both from the fact
that there is a poor quantitative agreement between the PDE
results and first order perturbation analysis in the case of
$m=2$, and from the fact that for $m=3$ it fails to predict the
existence of a drift velocity depending on an initial phase. The
reason for this discrepancy is that this analysis includes only
point-particle aspects of the soliton dynamics, ignoring
completely the internal structure of the soliton. In analogy with
the mechanism described in Ref. \cite{sq01}, one could expect a
strong contribution to transport coming from the soliton-phonon
interaction. Since this interaction arises only at second order in
a perturbation expansion, this explains why first order
calculations fail to  capture the phenomenon.

%

To elucidate the mechanism underlying soliton ratchets it is
useful to investigate the kink dc motion in more details. In Fig.
\ref{fig5} we depict the kink profile while  executing the ratchet
dynamics in the case of a $m=2$ driver. From this figure  the
existence of an internal oscillation (internal mode) of the kink
profile is clearly seen. Note that the internal mode is strongly
de-symmetrized with respect to the center of the kink . In the
panel (a) the kink moves from the left to the right and the mode
appears behind the kink. We checked that the motion of the kink is
locked to the external drive. This can be seen from Fig.
\ref{fig5}b, which shows the kink dynamics during two periods of
the driving force $E(t)$. This is in agreement with the results of
Ref. \cite{sq01}. The last panel, (c), shows the profile of the
kink while moving in opposite direction than the one in panel (a).
Note that when the kink moves to the left, the internal mode is on
the right side from the kink center, so it is again behind the
kink. This result indicates that there is an evident contribution
to the directed kink motion hidden in the asymmetric character of
the internal mode and in its interaction with the kink center of
mass. We also checked that by increasing the damping the internal
mode become smaller and smaller and almost disappears in the
over-damped limit. This correlates with the numerical results
showing the mean velocity of the kink rapidly decreasing with the
increasing of the damping.

It is worth to note that although the are no internal modes
frequencies in the spectrum of the small oscillation problem
around exact soliton solutions of the pure sine-Gordon equation,
these can arise from the perturbation field $\epsilon f$ when it
is switched on.
This makes the proposed internal mode mechanism for soliton
ratchets quite general.

%
Let us now briefly investigate the influence of a white noise on
the kink ratchet dynamics. In Fig. \ref{noise} we report the
contour plot for the kink motion in the case of a bi-harmonic
driver with $m=2$. We see that the noise introduces disturbances
on the profile but does not destroy the drift motion of the
soliton. We checked that this property remains true also if we
increase the amplitude of the noise up to the kink-anti-kink
nucleation limit.
Moreover, the existence of the phenomena in presence
of noise shows the validity of the above
mechanism also for the non-deterministic soliton ratchets.

\section{Boundary effects on soliton ratchets}

In this section we  discuss the effects of the system boundaries
on the ratchet dynamics. We have solved the problem for two types
of boundary conditions: free ends $u_x(0,t)=u_x(L,t)=0$ and
periodic boundary conditions $u(0,t)=u(L,t)+2 \pi n$, $n= \pm 1$,
where $L$ is the length of the sample. The behavior of kink and
anti-kink solutions does not differ for both cases except, of
course, for the behavior at the boundaries. For free boundaries
one can show by perturbation theory that the kink may  be
destroyed at the boundaries if its incoming kinetic energy is
below a certain threshold. This can be easily understood since at
the boundary the kink undergoes a large-amplitude oscillation
which, in presence of damping, expose the kink to larger
dissipation (the oscillation will be damped and the kink will not
be able to attain the $-2\pi$ value and get reflected as
anti-kink). The synchronization of the soliton motion with the
external ac field, as well as, the fact that kink and anti-kink
ratchet move is opposite directions, can allow sufficiently
energetic soliton ratchets to overcome reflections in presence of
damping. This is clearly shown in Fig. \ref{fig7}a where a
kink-anti-kink ratchet reflection is shown. The possibility to
overcome reflection, however, requires, to overcome a critical
energy threshold which depends on the system parameters. In Fig.
\ref{fig7}b we show the case in which the soliton ratchet is
destroyed at the boundary. At low damping the collision of the
kink with the boundary generate oscillations which decay, after
some time,  leading to the destruction of the kink (increasing the
damping the decay time quickly decreases).
%
The possibility of overcoming reflection depends also on the
relative phase of the driver and the internal mode oscillation on
top of the kink (this dependence can be tested by shifting the
initial incoming positions of the kink). In general, besides
kink-anti-kink reflections and kink destructions, more complicated
phenomena can arise, These including the possibility, for
particular values of damping, amplitudes and driver frequencies,
to sustain a standing breather at the boundary for very long time.
Similar phenomena were also observed in Ref.
\cite{lomdahl-samuelsen}.

To avoid the possibility of ratchet destruction at the boundaries
one can recourse to periodic boundary conditions. In this case
soliton ratchets, once established, will go on forever (we have
checked this numerically for very long computation times). This
opens the possibility of interesting physical applications as we
will discuss at the end of the next section.

\section{Conclusions}

In this paper we have considered a new way  to produce a directed
motion of a topological soliton  in presence of damping, by using
suitable ac drivers of zero mean. In particular we showed  the
possibility of establishing a ratchet dynamics for a kink of the
damped sine-Gordon equation when a periodic force, consisting of
two harmonic drivers, is applied. In contrast to previous works,
the observed dc motion does not requires any  asymmetry in the
potential of system, this making the phenomena easily accessible
to experimental situations.

We also showed that a first order perturbation analysis based on
collective coordinates does not provide a complete description of
the phenomenon.  The reason for this failure was ascribed to the
soliton-phonon interaction which, in a perturbation analysis,
appears only at second order. In particular we showed that the
soliton-phonon interaction manifests with the excitation of an
internal mode on the soliton profile which, in presence of
damping, can interact with its translation mode. This provides a
basic mechanism to convert the energy of the ac field into direct
motion for the soliton  which is valid for a wide class of
under-damped or moderately damped nonlinear systems. Numerical
simulations of the sine-Gordon equation confirm the validity of
the proposed mechanism.  We also investigated the influence of
boundary conditions on the ratchets dynamics in finite sine-Gordon
systems. For reflective boundaries we showed the possibility for
soliton ratchet to overcome reflections in presence of dissipation
(for periodic boundary conditions the soliton ratchet, once
established, will, obviously, go on forever). Finally, we showed
that the phenomenon of soliton ratchet is robust enough to
overcome the presence of the noise in the system.

This results open the possibility of interesting applications in
different fields. In the context of  Josephson junctions, for
example,  one can predict the existence of a zero field step (i.e.
steps in the current-voltage characteristic  related to the
resonant fluxon motion in the junction) in absence of dc bias
current and in presence of only bi-harmonic fields of zero
average. This effect should be best observable in annular
Josephson junctions where no boundary problems (kink destruction)
exist. We also remark that ``fluxon-ratchets'' inducing zero field
steps in Josephson junctions, have not yet been considered both
theoretically and experimentally (we will investigate this problem
in more details in a forthcoming paper).

Similar transport phenomena can be predicted also in a variety of
physical systems such as dislocations in crystals, spins waves in
magnetic chains, etc. Adjusting the relative phase of the drivers
one could control the direction and the velocity of these
excitations, this providing a way to control their dynamics. We
hope that the results of this paper will soon stimulate
experimental work in these directions.

\section{Acknowledgements}
It is a pleasure to thank P. L. Christiansen, J.C. Eilbeck, S.
Flach, Y.S. Kivshar, A. C. Scott, and M.R. Samuelsen for
interesting discussions.  MS wishes to thank the MURST for partial
financial support through the PRIN-2000 Initiative, and the
Department of Physics of the Technical University of Denmark, for
a two month Visiting Professorship during which this work was
started. YZ acknowledges the hospitality received at the Physics
Department of the University of Salerno, where part of this work
was done. Both authors acknowledge partial financial support from
the European Union grant LOCNET project no. HPRN-CT-1999-00163.


\newpage
Figure captions

%
\begin{figure}
\caption{Profiles of the bi-harmonic driver for
$m=2$ (continuous line) and $m=3$ (dashed line) for parameter
values $E_1=0.4$, $E_2=0.26$, $\omega=0.25$, $\theta=0$. For a
better comparison, a time shift of $4.548$ and of $-2\pi$ was
respectively applied to $m=2$ and $m= 3$ cases.} \label{fig0ms}
\end{figure}

%
\begin{figure}
\caption{Time dependence of the velocity $v(t)$ of
the kink center of mass (thick dashed curve). The continuous line
denote the force profile, while the dotted-dashed and dashed curves
represent the response to the single dark and light pulses,
respectively. The parameters are the same as in Fig. \ref{fig0ms}
with $m=2$ and $\alpha=0.15$.}
\label{fig3ms}
\end{figure}


%
\begin{figure}
\caption{Trajectories of the kink center of mass
derived from Eq. (\ref{6}) for the cases $m=2$ (continuous curve)
and $m=3$ (dashed curve). The other parameters are: $\omega=0.25$,
$\alpha=0.1$, $E_1=0.4$, $E_2=0.26$ and $\theta=1.61$.}
\label{fig1ms}
\end{figure}

%
\begin{figure}
\caption{Average velocity $\langle v \rangle$ of the
kink center of mass versus the the relative phase $\theta$ for the
same parameters as in in Fig. \ref{fig1ms}. The continuous curve
refers to the approximating function $\langle v \rangle= 0.058
\sin (\theta - 0.8)$.}
\label{fig2ms}
\end{figure}

%
\begin{figure}
\caption{ Contour plots of the velocity field
$u_x(x,t)$ with $\alpha=0.12$, $\omega=0.25$, $\theta=1.61$,  in
the cases (a)$E_1=0.4$, $E_2=0.26$ and (b)$E_1=0.66$, $E_2=0$.}
\label{fig1}
\end{figure}

%
%
\begin{figure}
\caption{ Contour plot of the velocity field
$u_x(x,t)$ of the anti-kink motion. System parameters are as for
Fig. \ref{fig1}a.}
\label{fig2}
\end{figure}

%
\begin{figure}
\caption{ The averaged kink velocity as a function
of damping constant $\alpha$ with $E_2/E_1=0.65$ for different
system parameters.
\\
Panel (a): $E_1=0.4$, $\theta=1.61$; $\omega=0.11$ ($\Box$),
$\omega=0.25$ ($\diamond$),
$\omega=0.35$ ($\bullet$) and
$\omega=0.65$ ($\circ$).\\
Panel (b):  $\omega=0.35$, $\theta=1.61$; $E_1=0.5$ ($\ast$),
$E_1=0.4$ ($\bullet$),
$E_1=0.38$ ($\Box$), $E_1=0.3$ ($\diamond$),
$E_1=0.2$ ($+$) and $E_1=0.1$ ($\circ$).\\
Panel (c): $\omega=0.25$, $E_1=0.4$; $\theta=0$ ($\circ$),
$\theta=0.8$ ($\diamond$),
$\theta=1.2$ ($\Box$) and $\theta=\pi$ ($\ast$). \\
The dashed lines show results obtained from numerical solution
of Eq. (\ref{6}) for $\omega=0.11$ and $\omega=0.25$ in (a);
for $E_1=0.4$ in (b) and for $\theta=0$ in (c).
\\
 The inset shows the dependence of the average
 velocity on the ``delay angle'' $\theta$ for
 $\alpha=0.15$ and the rest of parameters as in Fig. \ref{fig3}c.
 The dashed curve in the inset shows the fitting curve (see text
 for details).
}
\label{fig3}
\end{figure}

%
\begin{figure}
\caption{Dependence of the averaged velocity
$\langle v \rangle$ on damping for parameters as in Fig.
\ref{fig3}c but with different amplitudes: $E_1=0.4$ ($*$),
$E_1=0.45$ ($\circ$) and $E_1=0.5$ ($\diamond$). The ratio
$E_2/E_1$ has been kept constant, $E_2/E_1=0.65$.} 
\label{fig4}
\end{figure}

%
\begin{figure}
\caption{ The averaged kink velocity as a function
of the driver amplitude $E_1$
with $E_2/E_1=0.65$, $\theta=1.61$. The rest of the parameters were fixed as:\\
$\omega=0.11$, $\alpha=0.15$($*$); $\omega=0.35$, $\alpha=0.15$
($\diamond$); $\omega=0.35$, $\alpha=0.1$ ($\circ$);
$\omega=0.35$, $\alpha=0.05$ ($\Box$). The inset shows the log-log
dependence. } 
\label{fig10}
\end{figure}

%
\begin{figure}
\caption{Contour plot of the velocity field
$u_x(x,t)$ for a sine-Gordon kink driven by a bi-harmonic force
with $m=3$. System parameters are as for Fig. \ref{fig1}a except
$\alpha=0.15$ and $\theta=\pi$. } \label{k3}
\end{figure}

%
\begin{figure}
\caption{Kink profile and dynamics for
$\alpha=0.15$, $\omega=0.35$, $E_1=0.5$, $E_2=0.325$,
$\theta=1.61$ (a,b) and $\theta=1.61-\pi=-1.53$ (c). The profile
has been computed at the time moment $t=600$. The dashed line
shows the profile of the kink after one period of the external
drive.} 
\label{fig5}
\end{figure}

%
\begin{figure}
\caption {Countour plot for the displacement field
$u(x,t)$ for parameters as in Fig. 1a and noise amplitude
$D=0.1$.} 
\label{noise}
\end{figure}

%
\begin{figure}
\caption{Contour plots of kink (a) reflection and
(b) annihilation at the boundary. the system parameters are
$\omega=0.11$, $E_1=0.4$, $E_2=0.26$, $\theta=1.61$. The damping
parameter was $\alpha=0.08$ for the case (a) and $\alpha=0.12$ for
the case (b). The inset shows the anti-kink profile after
reflection at time $t=3200$. } 
\label{fig7}
\end{figure}

\end{document}